\documentstyle[11pt,epsf,newpasp,twoside]{article}
\def\msun{M_\odot}
\def\nbody{$N$-body\ }

\def\edcomment#1{\iffalse\marginpar{\raggedright\sl#1\/}\else\relax\fi}
\reversemarginpar

\begin{document}

\title{Two collaborative experiments in star cluster evolution}
 \author{Douglas C. Heggie}
\affil{University of Edinburgh, Department of Mathematics and
 Statistics, King's Buildings, Edinburgh EH9 3JZ, UK}

 \begin{abstract}
We summarise the specification and some previously unpublished results
of the collaborative experiment (``KyotoI'') that was carried out at the time of
the Kyoto General Assembly of the IAU.  As the subject has advanced
considerably since then, and codes have become more elaborate, we describe
an agreed proposal for a new experiment (``KyotoII'').
 \end{abstract}

 \section{Introduction}

Results from the very first collaborative experiment, with $N=25$
stars, were actually presented, not in 1997 in Kyoto, but thirty years
earlier at a meeting in Paris in 1967 (Lecar 1968).  Calculations for the
Kyoto experiment (which will here be referred to as KyotoI) took place
mainly in the run-up to the Kyoto General Assembly of the IAU in 1997,
where the results were presented.  It involved not only \nbody
integrations (up to $N = 65536$!) but also various kinds of
Fokker-Planck and even gaseous models (Heggie et al 1998).  Some
hitherto unpublished results are presented in the next section, along
with a summary of its specification.

KyotoI was restricted to pure stellar dynamics without primordial
binaries.  As the subject extends into the realm of ``all-inclusive'',
more realistic simulations, including the presence of primordial
binaries and the simulation of stellar evolution, it is timely to
initiate a more advanced collaboration.  In section 3 we present and
discuss the specification of a new problem which was largely agreed
during IAU Symposium 208 in Tokyo, though it will be referred to as
KyotoII.  Computations are only just beginning at the time of writing,
and results will be presented elsewhere.

\section{``First'' collaborative experiment (KyotoI)}

During the run-up to the IAU General Assembly in Kyoto, i.e. the
period May-August 1997, several groups around the world participated
in a collaborative experiment in star cluster evolution.  The aim was
to study the evolution of the same cluster by several
different techniques.  The initial specification of the cluster was as
follows:
\begin{enumerate}
\item  King Model, $W_0 = 3$ (non-rotating),  $r_t = 30$pc,  $M = 6\times10^4\msun$;
\item mass function $n(m)dm \propto m^{-2.35} dm$ for $0.1\msun < m < 1.5\msun$  ;
\item     No (primordial) binaries, no mass segregation;
\item     Tidal boundary conditions corresponding to circular motion round a point-mass galaxy;
\item       Heating by 3-body binaries;
\item       No stellar evolution, no collisions. 
\end{enumerate}

The 1997 collaborative experiment brought a number of benefits.  In
particular, new stellar dynamics was learned, because of problems
raised by the $N$-body results (see Baumgardt 2001), and the
specification has become a kind of benchmark problem which has been
occasionally used since (e.g. Giersz 2001).  Though only one
publication has emerged so far, a fault for which the author bears
total responsibility, considerable amounts of data became available
almost immediately on the web
(http://www.maths.ed.ac.uk/$\sim$douglas/experiment.html).  One drawback of the
specification also emerged.  Data had been requested corresponding to
times when the system had lost half of its mass (by escape) and at the
time of core collapse.  Unfortunately these times turned out to be
rather similar.

\begin{figure}
\plotfiddle{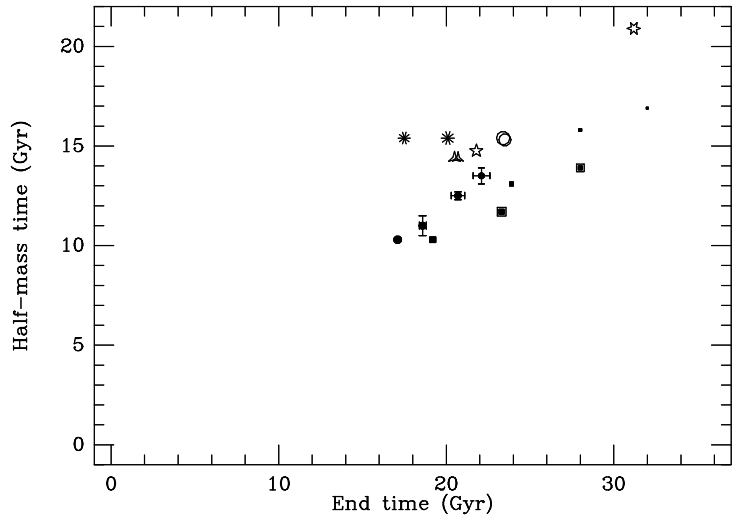}{2.6truein}{-90}{40}{40}{-175}{215}
\caption{``First'' collaborative experiment:  time at which half of
the mass had escaped, against end time.  
The origin is included to give an impression of
the relative variation in the results.  Some \nbody computations
include data from several realisations of the specification, and then
error bars are given.  In general, the size of the symbol relates to
some aspect of the reliability of the model, e.g. the number of mass
bins used in a Fokker-Planck simulation.  In some cases,
especially with the Monte Carlo code, a simulation can continue for a
lengthy period with almost negligible mass, and the end time gives a
poor indication of its reliability.  The fastest evolving \nbody
models are those with largest $N$.}
\end{figure}

\begin{figure}
\plotfiddle{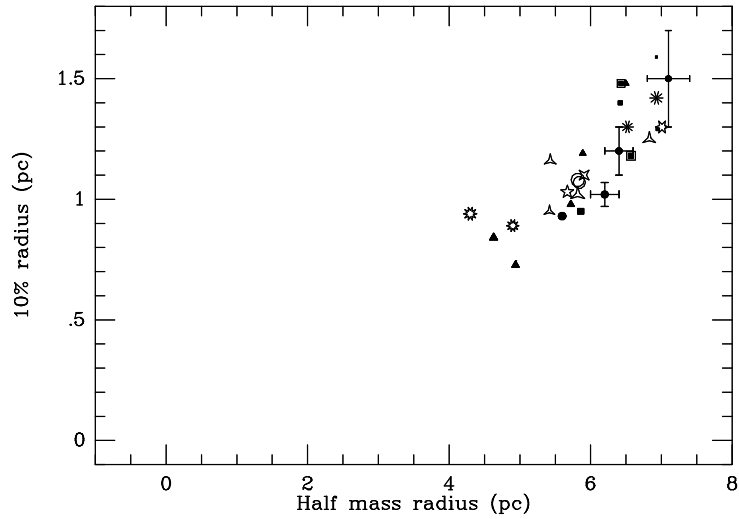}{2.6truein}{-90}{40}{40}{-175}{215}
\caption{``First'' collaborative experiment.  Half-mass and 10\%
Lagrangian radii at the time of core bounce.}
\end{figure}

\begin{figure}
\plotfiddle{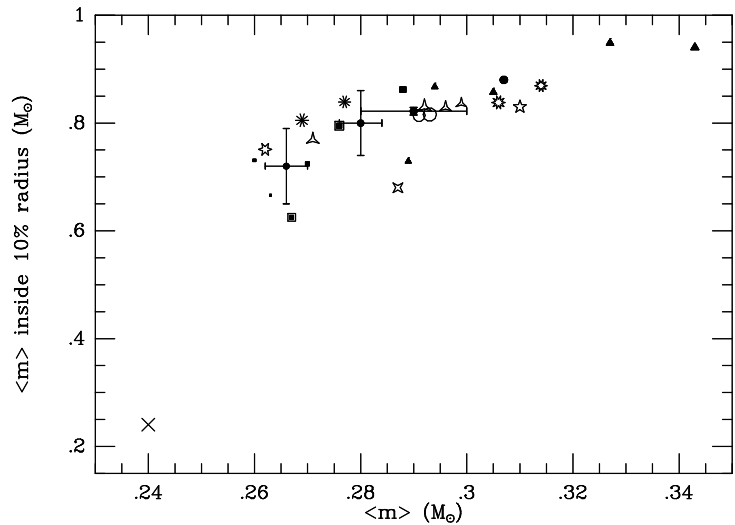}{2.6truein}{-90}{40}{40}{-175}{215}
\caption{``First'' collaborative experiment.  Global mean stellar mass,
and mean mass within the
10\%
Lagrangian radius, at the time of core bounce.  The cross at lower
left gives the initial values.  For a power law mass function, the
spread in abscissa values at core bounce corresponds to a variation in
the power law index from about $2.2$ to $1.7$; for the ordinates the
range is from about $-0.7$ to $+0.7$.}
\end{figure}

\begin{figure}
\plotfiddle{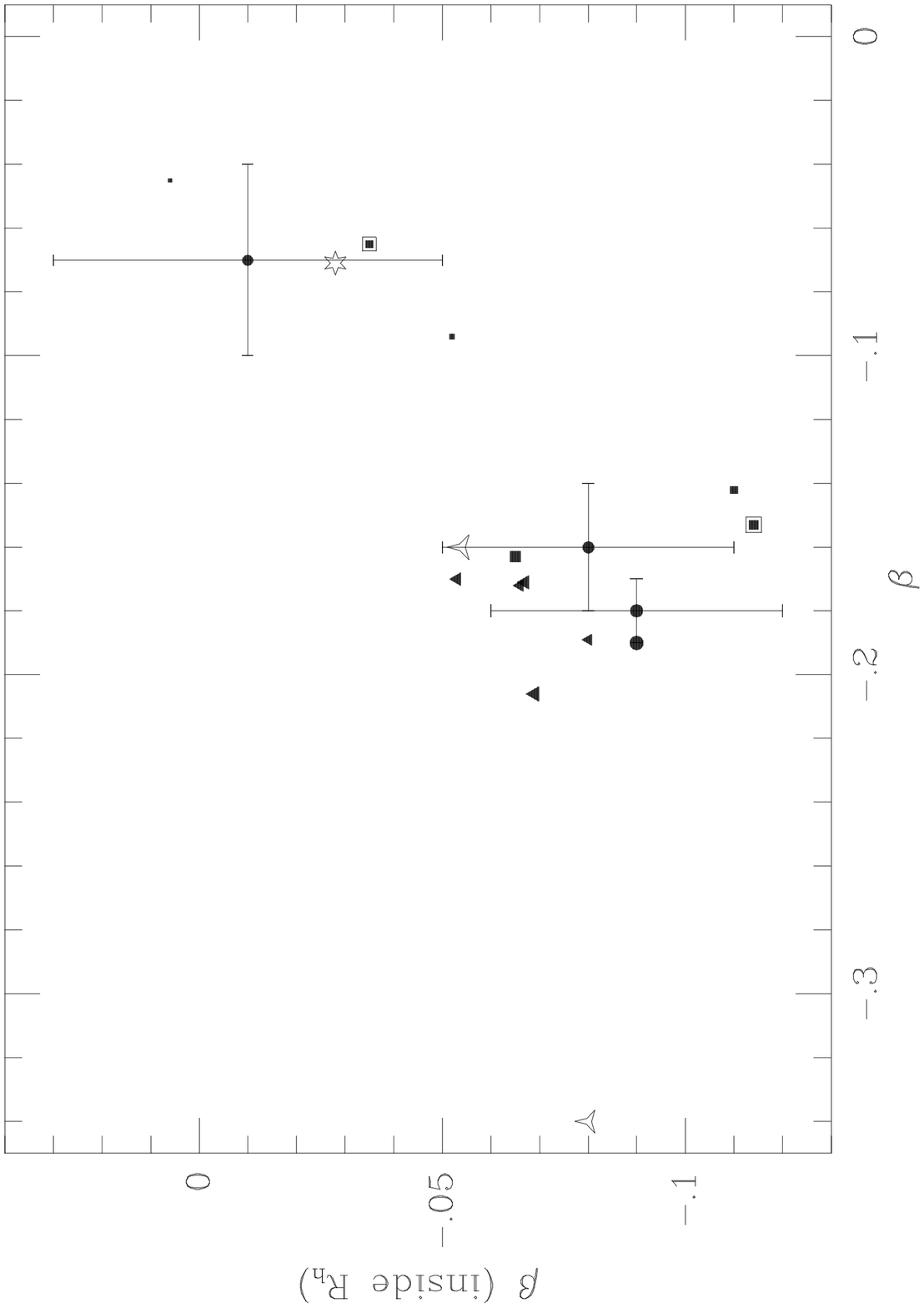}{2.6truein}{-90}{40}{40}{-175}{215}
\caption{``First'' collaborative experiment.  Global anisotropy $\beta$,
and anisotropy  within the half-mass radius, at the time of core
bounce.  $\beta$ is defined to be $1-\langle v_t^2\rangle/(2\langle
v_r^2\rangle)$, where $v_r$ and $v_t$ are radial and transverse
velocities, and the averages are {\sl not} mass weighted.}
\end{figure}

\begin{figure}
\plotfiddle{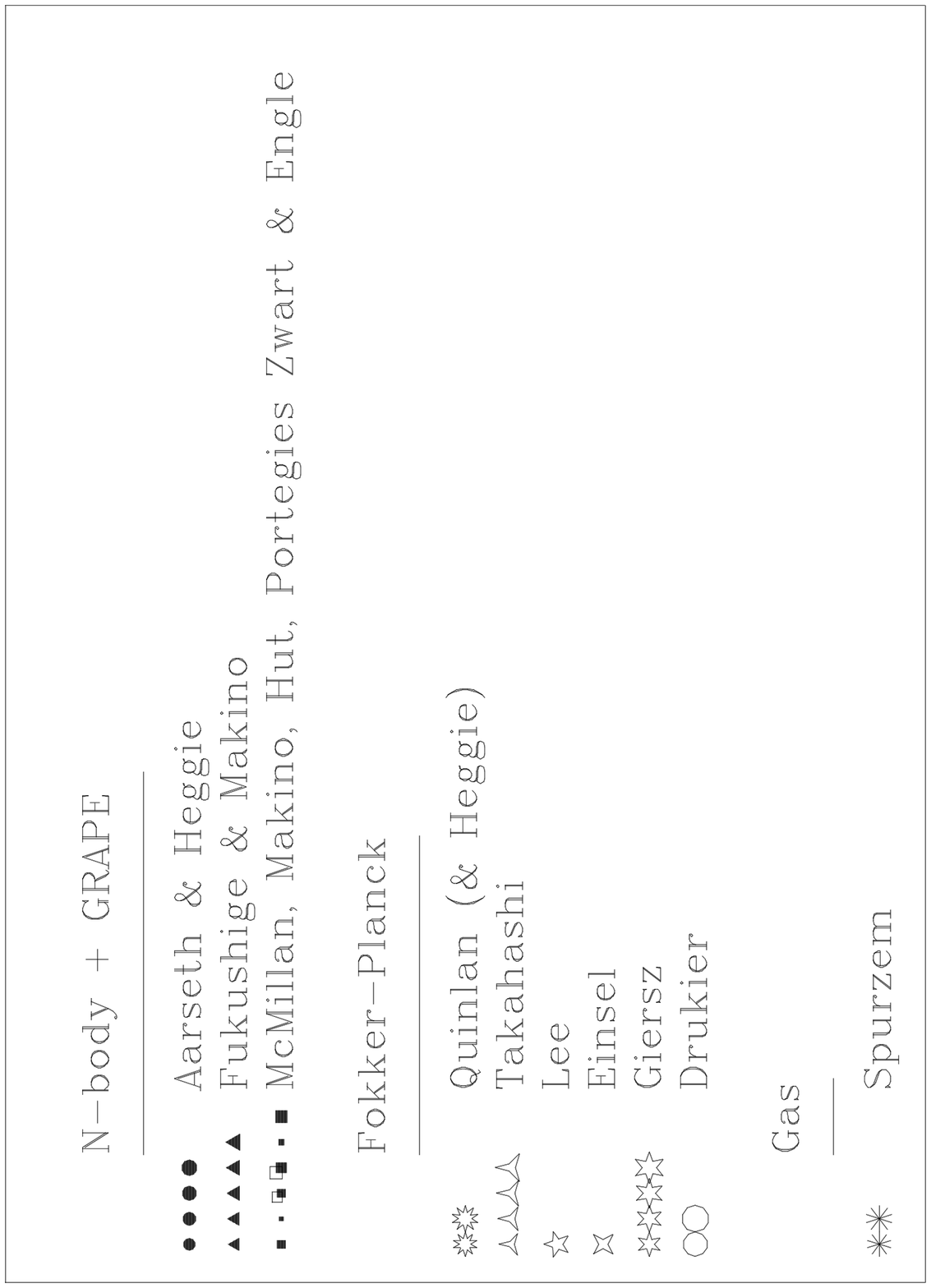}{2.2truein}{-90}{40}{40}{-175}{200}
\caption{``First'' collaborative experiment.  Key to the preceding figures.}
\end{figure}

Figs 1-4 illustrate some of the data which emerged from this
experiment.  Other data can be found in Heggie et al 1998.  The key to
the symbols is given in Fig.5, which also lists the persons who
participated.  In studying these figures it must be recalled that some
improvements have been made in these codes since the time of the
collaborative experiment, most notably in the treatment of boundary
conditions in Fokker-Planck codes (Takahashi \& Portegies Zwart 1998, 2000).

\section{   A new proposal}

\subsection{Specification}
Now that \nbody simulations in stellar dynamics are attempting a new
level of realism, it is timely to build on the previous collaborative
experiment by devising a more ambitious specification.  The following
specification had been under discussion for some time before the time
of the symposium, and was largely firmed up then.  The aim has been to
devise a model which displays an interesting interplay of stellar and
dynamical evolution, and is doable by as many groups as possible.  The
simulation of the recent evolution of M67 by Hurley et al (2001) has
been a very useful guide.  Their model, with $N_s = 5000$ single stars
and $N_b = 5000$ binaries (i.e. 50\% primordial binaries and $N = 15 000$ stars altogether) took
about 1 month on a GRAPE 4.

The agreed specification is as follows:
\begin{itemize}
\item   Initial structure
\begin{enumerate}
\item[1.] King (1966) model, $W_0 =5$
\item[2.] no initial mass segregation
\end{enumerate}
\item Boundary condition
\begin{enumerate}
\item[3.] tidal radius $r_t = 27.9$pc (initially)
\item[4.] tidal cutoff (i.e. instantaneous removal at $r_t$)
\end{enumerate}
\item Single stars
\begin{enumerate}
\item[5.] $N_s = 12288$
\item[6.] IMF: KTG3 (Kroupa, Tout \& Gilmore 1993\footnote{To be precise,
their eq.(14), using entries in Table
10 with $\alpha_1=1.3$}), restricted to $0.1\msun<m< 100\msun$
\end{enumerate}
\item Primordial binaries
\begin{enumerate}
\item[7.] $N_b = 4096$
\item[8.] eccentricity distribution $f(e) = 2e$
\item[9.] IMF for total binary mass: KTG1 (see Hurley et al 2001, eq.(1)),
restricted to  $0.2\msun<m<200\msun$
\item[10.] mass ratio $q<1$, uniform in the range giving component masses
in the range $0.1\msun<m< 100\msun$
\item[11.] $a$: uniform in $\log a$ in a range such that $a(1-e) > 2(R_1+R_2)$
and binding energy exceeds $10kT$\footnote{Stellar radii $R_i$ are from Eggleton, Fitchett \&
Tout 1989, and $1.5kT$ is the mean kinetic energy of single stars and
the barycentres of binaries}
\end{enumerate}
\item Stellar evolution
\begin{enumerate}
\item[12.] $Z =$ solar
\end{enumerate}
\end{itemize}

\subsection{Discussion}
Initial conditions, suitable for \nbody computations, have been
constructed in accordance with the above specification\footnote{See
the web page http://www.maths.ed.ac.uk/$\sim$douglas/kyotoII.html}.  It was
decided to do this once for all, and not to invite each participant to
create his own set of initial conditions.  The reason for this is that
different initial masses will lead directly to binaries with very different
histories, whereas use of a common set of initial conditions may allow
the fate of the same binary in different simulations to be compared.
The distributions of $q$ and $\ln a$ in the initial conditions are
not, of course, uniform, as the ranges of $q$ and $a$ depend,
respectively, on the
total mass of a binary and the masses and radii of the components.

It follows from the above specification that the initial proportion of
primordial binaries is $f_b = 25\%$, and that the total initial number
of stars is $N = N_s + 2N_b = 20480$.  The initial conditions which have been
constructed lead to a total initial mass
$M = 10174\msun$ approximately.  For a point-mass galaxy, such a mass
yields the required tidal radius at a galactocentric distance of
$8.5$kpc if the circular speed is $220$km/s.

Fig 6 provides a summary of the kinds of dynamical and other
processes which the two collaborative experiments involve.  It also
gives an indication of the kinds of computer code to which the
specified problems are accessible.  Even the new proposed experiment
does not test certain aspects of the dynamics:  unsteady tides, for
example (as for a cluster on an elliptical galactic orbit, or with
disk shocking), or rotation (cf. Kim et al 2001).

\begin{figure}
\plotfiddle{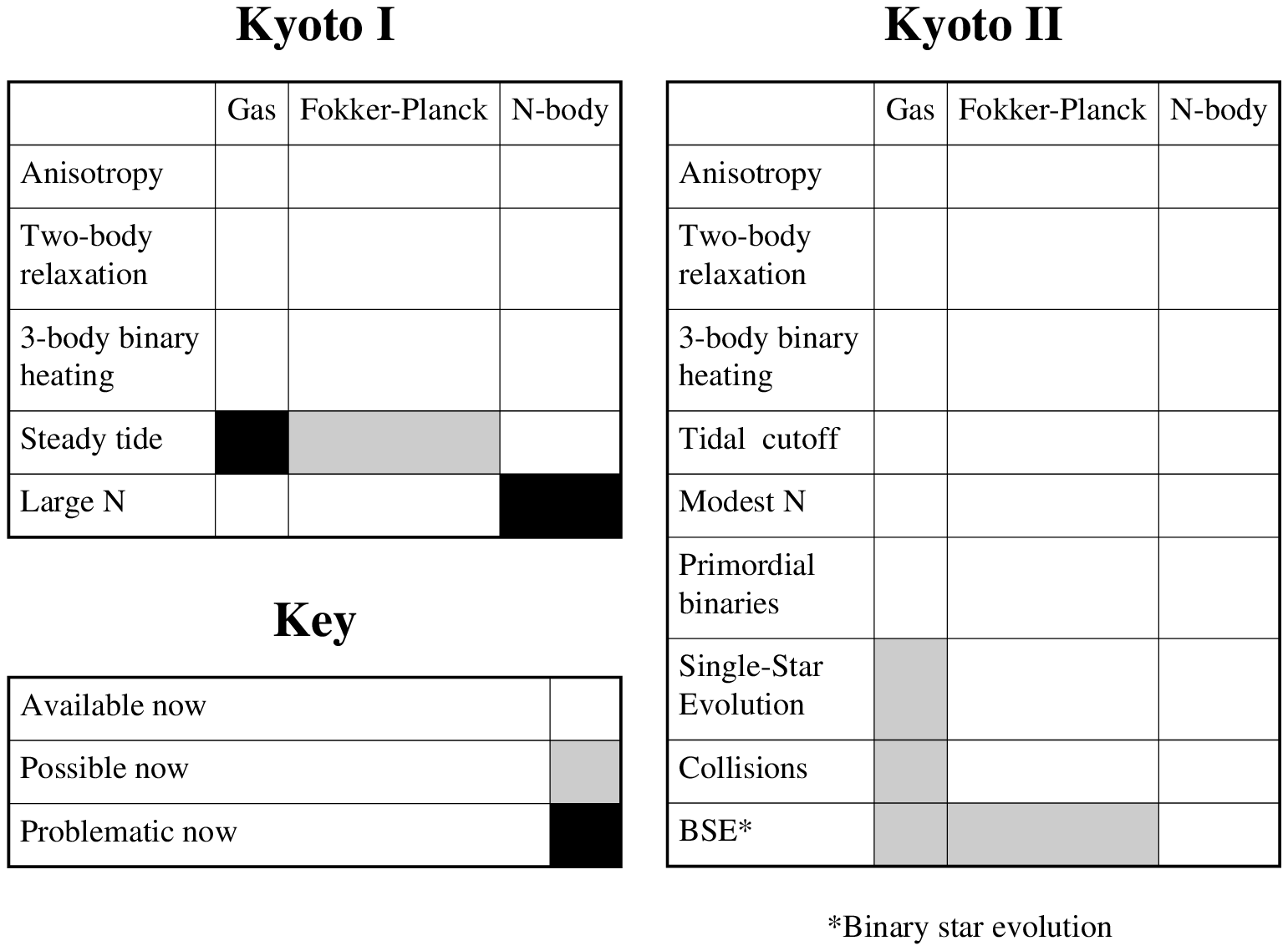}{2.7truein}{0}{75}{75}{-220}{-300}
\caption{The dynamical and other features of the two collaborative
experiments.  ``Available now'' means that the author is aware of the
existence of at least one version of this kind of code where the stated
feature is implemented.  ``Possible now'' means that it could be
implemented within the time scale of the current experiment, but not
that this would necessarily be straightforward or easy.  ``Problematic
now'' means that there are unsolved difficulties.}
\end{figure}

Notice the choice of a tidal {\sl cutoff}, rather than the more realistic
tidal field adopted in the ``first'' collaborative experiment.  The reason
for this is that it has become clear that implementation of tidal
boundary conditions in the Fokker-Planck method is a delicate issue,
and it is in the context of a tidal cutoff that the problems have been
most thoroughly studied (Takahashi \& Portegies Zwart 1998, 2000).  

In much
the same way, the value of $N$ here is much less relevant to the
dynamical evolution of globular clusters than the value selected in
the ``first'' experiment.  One  reason is that it is known to be
difficult to scale $N$-body results to large enough $N$, though the
problems are minimised by the use of a tidal cutoff (Baumgardt 2001).
In principle, the collision cross section can be scaled (Portegies
Zwart et al 1999) but, more importantly, there is no experience in the scaling (with $N$) of \nbody
models which include stellar evolution. 

\subsection{   Output}

According to results of Vesperini \& Heggie (1997) and Portegies Zwart
et al (2001b) (see Heggie, this vol.), if binaries were treated as
single stars with the combined mass of both components, and if stellar
evolution were neglected, the lifetime in a tidal {\sl field} would be
$4.0$Gyr.  Nevertheless, an unconfirmed computation with ``inert''
binaries (see below) implies that the lifetime within a tidal cutoff
is about $2$Gyr.  Therefore an output interval of $1$Gyr would be too
small.  In fact the specification of the collaborative
experiment requires output at intervals of $0.5$Gyr (at least), and the minimal required
output is
\begin{enumerate}
\item     Mass $M$
\item	Half-mass radius     $r_h$ 
\item $N_s$, $N_b$
\item Number of blue stragglers, $N_{bs}$, and degenerate stars,
$N_{deg}$
\item Luminosity      $L$
\end{enumerate}
Optional  additional output includes
\begin{enumerate}
\item   Luminosity function
\item   Mass function
\item   Colour-magnitude diagram
\end{enumerate}
and no doubt more.

\subsection{   Partial calculations}

The issue of stellar evolution is left largely unspecified in the
above list, except for the metallicity, $Z$.  Many codes which might be
used for the study of the evolution of globular clusters lack all but
the most rudimentary treatment of stellar evolution.  Appropriate
sources of formulae on single star evolution are Eggleton, Fitchett \&
Tout (1989) and Hurley, Pols \& Tout (2000).  Evolution of binary
stars is even more involved, but suitable treatments are described in
Tout, Aarseth \& Pols (1997) and Portegies Zwart et al (2001a), and in
the references in the latter.

In principal, stellar evolution, where it is lacking, could be
implemented in several codes within the time scale of the
collaborative experiment.  Even so, it is very likely that the number
of participants in the full calculation will be very small.  But there
is also a set of restricted but useful computations for which results could be
achieved by many more groups.  These are
\begin{enumerate}
\item {\sl Stellar evolution mass loss only}  This approach, adopted
by Chernoff \& Weinberg (1990), may be updated through use of the
formulae of Eggleton et al (1989) for the evolution time.  At this
time, the mass of each star is replaced by the appropriate remnant
mass.
\item {\sl    No stellar evolution}  In this case the evolution is
restricted to pure dynamical evolution, as in the ``first'' collaborative
experiment, though now including primordial binaries.
\item {\sl Inert binaries} Here, even the internal degrees of freedom
of the binaries are ignored.  Suitable \nbody initial conditions are also
available from the web page.
\end{enumerate}

\subsection{   Timetable}

The idea of a second collaborative experiment arose in the spring of
2001, and  Table 1 summarises the time scale to which the
collaborative experiment should work.

\begin{table}
\begin{center}
\caption{Timetable of collaborative experiment}
\end{center}
\begin{tabular}{rl}
\tableline
Until end August 2001 &    Discussion of specification \\
Until end February 2002 &    ``Blind'' calculations (see text) \\
February 2002	& Publication of data, followed by  revised 
   calculations \\
June 2002	& First discussion of results\\
\tableline\tableline
\end{tabular}
\end{table}

During the period when ``blind'' calculations are carried out, the
only results which will be made available are a specification of the
kind of calculation that has been carried out (e.g. ``inert binaries,
Fokker-Planck'').  At the end of February, 2002, the results obtained
so far will be made available on the web page.  This may reveal
shortcomings or disagreements in the calculations which have been
made, and so there will be an opportunity for revised calculations to
be performed before the results are summarised publicly.  This initial
public account of the results may be given at a workshop on the topic
{\sl Integrating Stellar Evolution and Stellar Dynamics}, June 17-21,
2002, at the American Museum of Natural History.  In due course a
definitive paper will be prepared, discussing the results in detail.

\acknowledgements

I thank the organisers for their support, and Piet Hut, who originally
suggested the idea of a second collaborative experiment.  I am
grateful to colleagues from the first experiment, S.J. Aarseth,
G. Drukier, T. Fukushige,
M. Giersz, H.M. Lee, R. Spurzem and K. Takahashi for permission to include a
summary of the results in this paper.  I apologise to other
participants whom I was unable to contact before the deadline for submission.

\end{document}